# An Exploration on Brain Computer Interface and Its Recent Trends

T. Kameswara Rao
Assoc. Prof., Dept of CSE
Visvodaya Technical Academy
Kavali, AP, India

M. Rajyalakshmi
Assoc. Prof., Dept of CSE
Visvodaya Technical Academy
Kavali, AP, India

Dr. T. V. Prasad
Dean of Computing Sciences
Visvodaya Technical Academy
Kavali, AP, India

*Abstract—* **Detailed exploration on Brain Computer Interface (BCI) and its recent trends has been done in this paper. Work is being done to identify objects, images, videos and their color compositions. Efforts are on the way in understanding speech, words, emotions, feelings and moods. When humans watch the surrounding environment, visual data is processed by the brain, and it is possible to reconstruct the same on the screen with some appreciable accuracy by analyzing the physiological data. This data is acquired by using one of the non-invasive techniques like electroencephalography (EEG) in BCI. The acquired signal is to be translated to produce the image on to the screen. This paper also lays suitable directions for future work.**

## I. INTRODUCTION

In general, all living beings can connect with the surrounding environment with their five senses, but major role is played by the visual data, which is percepted by eyes. Vision begins with light passing through the cornea. All the living beings can see the surroundings, along with some animals and birds, humans also can identify the colors due to their eye structure. Unlike some other living beings, human eye contains three types of cones (so named because of their shape) that are sensitive to red, green and blue colors, and identify the colors under suitable lighting conditions.

Rods (so named because of their shape) also are the part of the eye structure along with the cones, and are involved in identifying objects under dim lighting conditions because they are more sensitive to dim light than cones and do not sense the color, produce grey scale data of the objects. Rods are common in all living beings' eye structure. Figure 1 shows the arrangement of rods and cones. The human eye's 125 million visual receptors (composed of rods and cones) turns light into electrical signals and send to the brain through the optical nerve [1].

## II. WORKING OF BRAIN

Brain is completely covered by a network of different types of neurons, which are the processing units of the data. Figure 2 shows the different types of neurons according to their structure. Most neurons have four functionalities in common, viz., input, trigger, conductile and output. These neurons percepts the data from synaptic terminals and it is processed in the cell body. Neurons can perform two types of actions namely fire and inhibit. They fire when the severity of the data is more than the threshold value which is set based on the experience/ training or they inhibit if severity is below the threshold value [1].

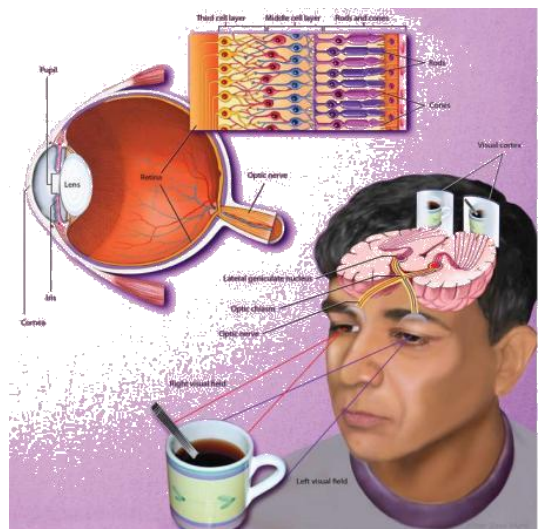

Figure 1 Identification of objects and their colors by brain

Brain is assumed to be divided into 52 discrete local areas by Korbinian Brodmann, and named as cyto-architectural map [3], each is processing specific type of data. For example visual data is gathered to visual cortex areas 17, 18, 19. These areas are depicted in Figure 3.

When the visual data is processed by the neurons which are in the areas 17, 18 and 19 [3], there, the neurons of that area, generate electric pulses or signals and magnetic fields [2] to perform actions by actuators of the body. Usually, the electric signals of the brain are generated by pumping the ions like sodium (Na+), potassium (K+), calcium (Ca++), chlorine (Cl−), through the neuron membranes in the direction ruled by the membrane potential[14].

## III. PHASES IN BRAIN COMPUTER INTERFACING

There are many phases in Brain Computer Interfacing. The major phases are as follows:

1. Signal Acquisition
2. Signal Pre-Processing
3. Signal Classification
4. Computer Interaction

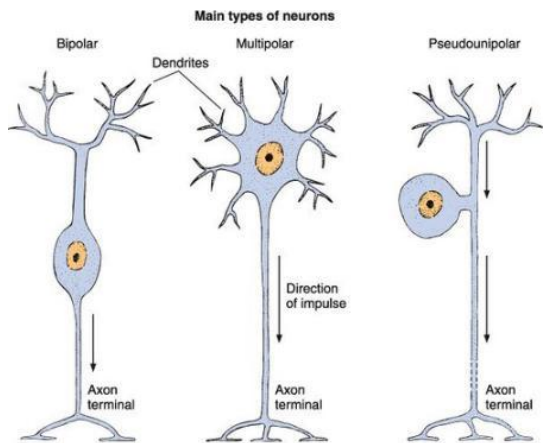

Figure 2 Structural types of Neurons

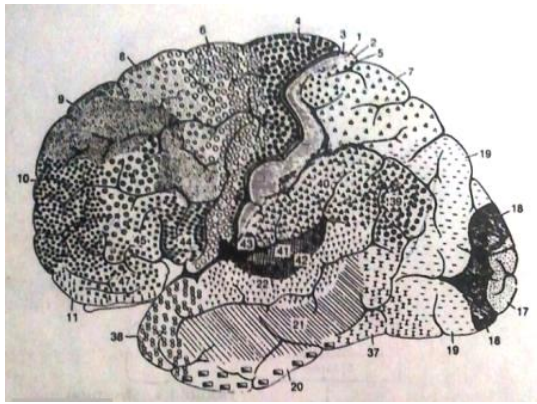

Figure 3 Cytoarchitectural map of cerebral cortex.

*A. Signal Acquisition:*

The electric signals generated by the neurons are acquired and processed by the signal acquisition and processing techniques and devices. In general, there are two types of brain signal acquisition techniques [4][7]:

 *a) Invasive acquisition*

These techniques are used to capture the brain signals (also referred as electrophysiological signals) using implanted electrodes in brain tissue directly from the cerebral cortex, as shown in the Figure 4.

This Invasive signal acquisition should require surgery to implant the sensors or electrodes. These electrodes are implanted by opening the skull through a proper surgical procedure called craniotomy [5]. The electrodes are placed on the cortex; the signals acquired from these electrodes are called electro-corticogram (ECoG) [5]. Invasive signal acquisition techniques give excellent quality of signals.

But, even though the Invasive signals, can provide fast and potentially information-rich information, there are drawbacks in this technique. Some of them are, Invasive technique must require a surgery which is an ethical controversy. That's why Invasive techniques are almost exclusively investigated in animal models. At the same time, the threshold for their use will be higher.

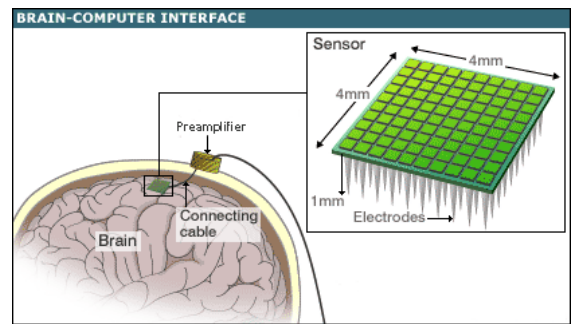

Figure: 4. Invasive data acquisition

Electrodes are piece of wire or metal pins, sometimes they may not work properly in the brain.

 *b) Non-Invasive acquisition*

These techniques are used to capture the signals or electrophysiological signals from the scalp, as seen in the Figure 5, by using the technologies like electroencephalogram (EEG), functional magnetic resonance imaging (fMRI), magneto-encephalogram (MEG), P-300 based BCI etc. [5]. Among non-invasive Brain Computer Interfaces (BCIs), electroencephalogram (EEG) has been the most commonly used for them because EEG is advantageous in terms of its simplicity and ease of use, which meets BCI specifications when considering practical use.

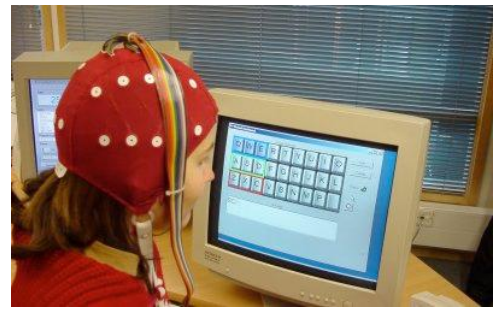

Figure: 5. Non-Invasive data Acquisition.

In general, EEG signals (EEGs) can be classified into two categories, spontaneous EEGs and stimulus evoked EEGs. Focusing on stimulus evoked EEG signals called P300 and Visual Evoked Potentials (VEPs) are often utilized for BCIs [17]. Both types of BCIs extract the intention of users. While P300 signals are thought to be derived from the thoughts of users, VEPs are simply derived from physical reaction to visual stimulation. In that sense, VEP-based BCIs are thus known as the simplest BCIs [19]. In this paper, only electroencephalogram is concentrated as the signal capturing technology.

These captured signals are used for BCI and are too weak, about 100 μV [6], that's why they have to be amplified. The capturing devices are wired or wireless. A wired device and a wireless device are shown in Figure 5 and 6 respectively.

The signals are amplified to 10000 times. A sample EEG wave is shown in Figure 7 along with a Frequency-Amplitude graph.

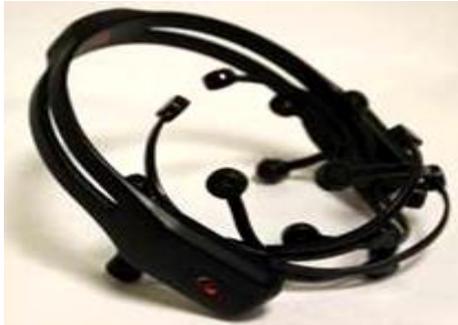

Figure: 6. A wireless non-invasive signal capturing device

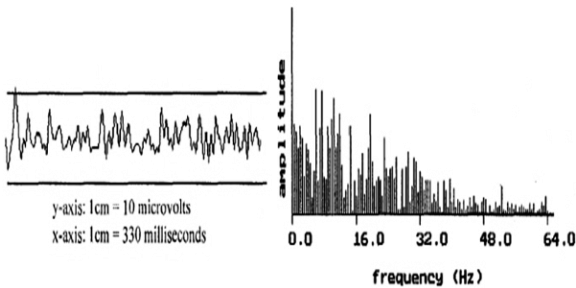

Figure 7. An EEG wave form recorded by a fore head electrode and its spectrum

*B. Signal Pre-Processing*

Whatever the technique we follow, the unwanted signal i.e., noise in the signal is inevitable. EEG recordings typically not only contain electrical signals from the brain, but also several unwanted signals like

- interference from electronic equipment, as for example the 50 or 60Hz, power supply signals,
- electromyography (EMG) signals evoked by muscular activity,
- ocular artifacts, due to eye movement or blinking.

Those unwanted components may bias the analysis of the EEG, and may lead to wrong conclusions. So the signal must be pre-processed to remove the noise. There are several preprocessing techniques to remove unwanted signals from EEG [18].

- **Basic Filtering:** The spurious 50 or 60Hz power supply signals are typically removed by a band-stop filter, which is a filter that passes most frequencies unaltered, but attenuates those in a specific range (e.g., at 50 or 60Hz) to very low levels.
  However, other artifacts such as electromyogram (EMG) signals and ocular artifacts typically affect a large frequency band and their spectrum may vary over time. Therefore, bandstop filters are usually not effective to eliminate such artifacts. One is often interested in specific frequency bands in the EEG, such as 4–8Hz (theta), 8–10Hz (alpha 1), 10–12Hz (alpha 2), 12–30Hz (beta), and 30–100Hz (gamma). Such frequency bands are usually extracted by a band-pass filter, which is a filter that passes frequencies within a certain range and rejects (attenuates) frequencies outside that range.
- **Adaptive Filtering:** The spectrum of artifacts is often a priori unknown. Therefore, applying a fixed filter to EEG data would not be effective to remove artifacts. The filter needs to adapt to the spectrum of the recorded EEG: it should attenuate the recorded EEG in frequency ranges that mostly contain artifacts. For instance, instead of using an online notch filter centered at a fixed frequency, one may apply an offline notch filter whose characteristics are determined by the spectrum of the recorded EEG. One may additionally use EOG (electro-oculography) or EMG (electromyography) measurements to design the adaptive filter, since those measurements are usually strongly correlated with artifacts.
- **Blind Source Separation:** An alternative approach, known as "blind source separation" (BSS), starts from the assumption that EEG signals can be described, to a good approximation, by a finite set of sources, located within the brain; each of those sources generate certain components of the EEG. Besides EEG, one sometimes also incorporates EOG and EMG signals into the analysis. In the context of artifact rejection, one makes the additional assumption that artifacts are generated by a subset of the extracted sources; one removes those sources, and next reconstructs the EEG from the remaining "clean" sources. Once the signals are acquired, it is inevitable to avoid noise. So the noise must be eliminated by using the algorithms like Adaptive filtering algorithm or RLS algorithm [13].

The recent technologies like merging of Translation Invariant Wavelet and ICA giving better results in noise filtering [15], shown in Figure 8.

*C. Signal Classification*

Since the brain signals or EEG are continuously captured by the capturing devices that have numerous electrodes, simultaneously capturing the signals in large amounts, it is not possible to clearly classify the waves. The waves are sometimes classified on both frequency and on their shape. There are six types of important signals [4], [8-12].

1) **Beta waves:** Frequency of these waves is between 13 and 30 Hz and the voltage or amplitude is very low about 5 to 30 µV. Beta waves are released when brain is active, in thinking, focusing on the problem solving etc. These waves' frequency can reach 50 Hz, during brain's intense activity. These waves are depicted in right side of the Figure 9.

2) **Alpha waves:** Frequency of these waves is between 8 and 13 Hz and the voltage or amplitude is about 30 to 50 µV. these waves are released when brain is relaxed or at inattention from the occipital and frontal cortexes. Alpha wave can reach a frequency of 20 Hz, which is a beta range, but has the characteristics of alpha state rather than beta. Alpha alone indicates a mindless state or empty state rather than relaxed of passive states. These waves are reduced by opening eyes or ears or by creating anxiety or tension mental concentration etc. These waves are depicted in left side of the Figure 9.

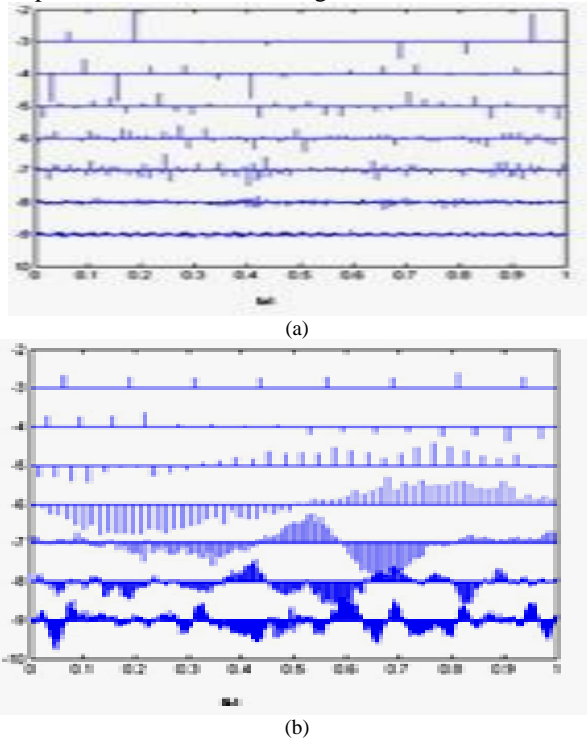

(a)

(b)

Figure 8 (a) wave coefficient before de-noising (b) wave coefficient after de-noising [14]

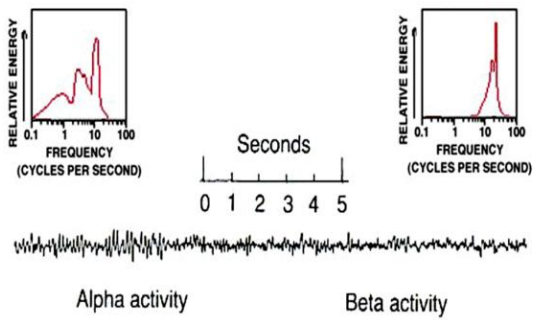

Figure 9. Alpha and Beta waves

3) **Theta waves:** Frequency of these waves is between 4 and 7 Hz and the voltage or amplitude is about 20 μV. Theta waves are generated when the brain is under emotional tensions, stress, frustration, disappointment etc. Theta waves are also released in unconsciousness or deep meditation. The peak frequency of theta waves is 7 Hz. These waves are depicted Figure 10.

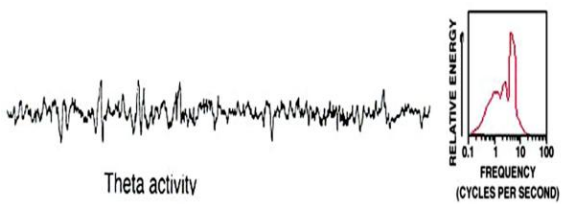

Figure 10. Theta waves

**Delta waves:** Frequency of these waves is between 0.5 and 4 Hz and the voltage or amplitude is varying. These waves are released deep sleep or when physical defects are there in the brain. These waves are depicted in Figure 11.

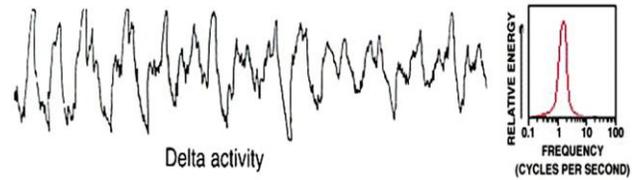

Figure 11. Delta waves

4) **Gamma waves:** Frequency of these waves is 35Hz or high. These waves are released when the brain is under a stream of consciousness.

5) **Mu waves:** Frequency of these waves is between 8 and 12 Hz. These waves are released with spontaneous nature of the brain like motor activities etc. These waves looks like alpha waves but alpha waves are recorded at occipital cortex and mu waves are recorded at motor cortex. These waves are depicted in Figure 12.

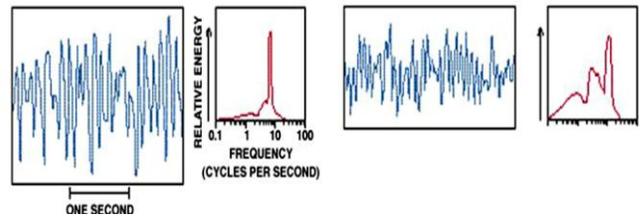

Figure 12. Mu and Alpha waves

Once the signals are cleaned, they will be processed and classified to find out which kind of mental task the subject is performing.

By using different BCI principles, the subjects participating in the P300 study had to spell a 5 character word with only 5minutes of training.

EEG data were acquired to train the system while the subject looked at Recent Advances in Brain-Computer Interface Systems, a 36 character matrix to spell the word WATER. During the real-time phase of the experiment, the subject spelled the word LUCAS.

For the P300 system 72.8 % were able to spell with 100 % accuracy and less than 3 % did not spell any character correctly as shown in Table 1.

Interesting is also that the Row-Column Speller reached a higher mean accuracy compared to the single character speller which produces higher P300 responses. This can be explained by the longer selection time per character for the SC speller [11].

*D. Computer Interaction*

Once the signals are classified, they will be used by an appropriate algorithm for the development of a certain application.

TABLE 1. CLASSIFICATION ACCURACY OF P300 EXPERIMENTS [11]

| Classification Accuracy [%] | Row-Column Speller: Percentage of sessions (N=81) | Single Character Speller: Percentage of Sessions (N=38) |
|---|---|---|
| 100 | 72.8 | 55.3 |
| 80-100 | 88.9 | 76.3 |
| 60-79 | 6.2 | 10.6 |
| 40-59 | 3.7 | 7.9 |
| 20-39 | 0.0 | 2.6 |
| 0-19 | 1.2 | 2.6 |
| Average Accuracy of all subjects | 91.0 | 82.0 |
| Mean of subjects f participated in RC and SC (N=19) | 85.3 | 77.9 |

## IV. LIMITATIONS

Some limitations in implementing the BCI system are variability in the acquired EEG signals. Different types of signals are acquired from the same person in different sessions, different signals are acquired when many people are performing the same mental task and EEG signals are affected by the person's eye blinks, muscular movements, suddenly hearing sound and interference from electronic devices etc [7].

Based on these different types of applications like, character recognition, phrase recognition, object movement etc., it is observed that it is also possible to reconstruct the images and videos on to the screen from brain signals [20], [16]. This is depicted in Figure 13 and 14.

When the human watches the environment, the brain starts analyzing the image or video. It generates the signals according to the situation and information it received.

All the time, it is not possible to reconstruct the accurate image or video on to the screen because of the limitations in BCI, but it can be obtained some abstract image on to the screen as the Figure 13 depicts.

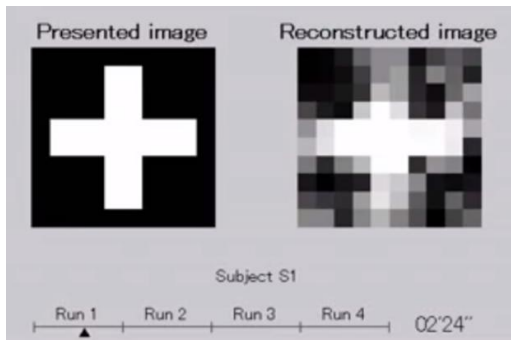

Figure 13 Binary image re-construction from brain signals [20].

This can be enhanced in such a way that, any color or object or image or video is seen, that can be produced on to the computer by properly capturing the signals of the brain. Not only binary images, but color images also reconstruct able.

Most of the binary image is reconstructed, but color image is not since the composition of colors in the brain is not as easy as binary.

So, to produce the color image or video on to the screen accurately from the brain signals, further much more great brain signal acquisition and analysis has to be taken place as future enhancement.

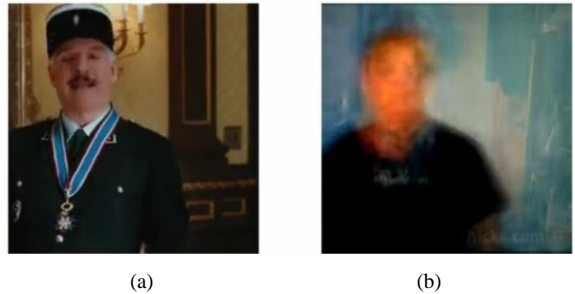

(a)  (b)

Figure 14. Color image reconstruction from brain signals [16]: a) Presented Clip b) Clip reconstructed from brain activity

## V. CONCLUSION AND FUTURE DIRECTIONS

Many complex processes and system would operate on the basis of thought in the future. Currently, the field of BCI is in infancy stage and would require deeper insights on how to capture the right signals and then process them suitably. The advancements are limited to recognition of certain words, expressions, moods, etc.

Efforts are being made to recognize the objects as they are seen by the brain. These efforts will bring in newer dimensions in the understanding of brain functioning, damage and repair. It is possible to recognize the thoughts of the human brain by capturing the right signals from the brain in future.

Research work can be taken up on the signal processing and analysis to tap the thoughts of the human. But present operating systems and interfaces are not suitable for working with thought based system. They have to be modified accordingly.

Processing and understanding thoughts for different purposes can lead to security and privacy issues. Ethical and standardization issues can also come up, which needs to be resolved earliest.

AUTHORS PROFILE

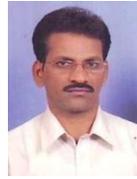
**Mr. T. Kameswara Rao** received his Masters in Computer Applications from University of Madras in 2004 and Masters of Engineering from Satyabhama University in 2007. He is currently associated with the Dept. of Comp. Sc. & Engg. at Visvodaya Technical Academy, Kavali, AP, India, as Associate Professor. He has over 7 years of teaching experience at under graduate and graduate level. His areas of interest are biometrics, brain computer interface, artificial intelligence, artificial neural networks, psychology etc.

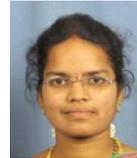
**Ms. M. Rajyalakshmi** received her Masters in Computer Science and Engineering from Jawaharlal Nehru Technological University, Anantapur in 2011. She is currently associated with the Dept. of Comp. Sc. & Engg. at Visvodaya Technical Academy, Kavali, AP, India, as Associate Professor. She has over 5 years of teaching experience at under graduate and graduate level. Her areas of interest are language processing, brain computer interface, automata theory, artificial neural networks, etc.

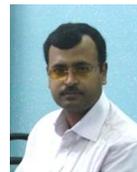
**Dr. T. V. Prasad** has over 17 years of experience in industry and academics. He has received his graduate and master's degree in Computer Science from Nagarjuna University, AP, India. He was with the Bureau of Indian Standards, New Delhi for 11 years as Scientist/Deputy Director. He earned PhD from Jamia Millia Islamia University, New Delhi in the area of computer sciences/ bioinformatics. He has worked as Head of the Department of Computer Science & Engineering, Dean of R&D and Industrial Consultancy and then as Dean of Academic Affairs at Lingaya's University, Faridabad. He is with Visvodaya Technical Academy, Kavali as Dean of Computing Sciences. He has lectured at various international and national forums on subjects related to computing. Prof. Prasad is a member of IEEE, IAENG, Computer Society of India (CSI), and life member of Indian Society of Remote Sensing (ISRS) and APBioNet. His research interests include bioinformatics, consciousness studies, artificial intelligence (natural language processing, swarm intelligence, robotics, BCI, knowledge representation and retrieval). He has over 73 papers in different journals and conferences, and has authored six books and two chapters.